\documentstyle[aasms4]{article}

\makeatletter
\let\@internalcite\cite
\def\cite{\@ifstar{\citeyear}{\citefull}}
\def\citefull{\def\citeauthoryear##1##2##3{##1, ##3}\@internalcite}
\def\citeyear{\def\citeauthoryear##1##2##3{##3}\@internalcite}

\def\@citex[#1]#2{\if@filesw\immediate\write\@auxout{\string\citation{#2}}\fi
  \def\@citea{}\@cite{\@for\@citeb:=#2\do
    {\@citea\def\@citea{; }\@ifundefined
       {b@\@citeb}{{\bf ?}\@warning
       {Citation `\@citeb' on page \thepage \space undefined}}%
{\csname b@\@citeb\endcsname}}}{#1}}

\def\@cite#1#2{(#1\if@tempswa , #2\fi)}
\def\@biblabel#1{}
\makeatother

\newcommand{\timehr}{\mbox{${}^{\rm{\scriptsize{h}}}$}}
\newcommand{\timemin}{\mbox{${}^{\rm{\scriptsize{m}}}$}}

\newcommand{\smass}{\mbox{$M_{\sun}$}}
\newcommand{\HII}{{\sc H\thinspace ii}}
\newcommand{\hii}{{\sc H\thinspace ii}}

\newcommand{\hi}{{\sc H\thinspace i}}
\newcommand{\COa}{$^{12}$CO $J=1\to0$}
\newcommand{\COb}{$^{12}$CO $J=2\to1$}
\newcommand{\tCOb}{$^{13}$CO $J=2\to1$}
\newcommand{\COc}{$^{12}$CO $J=3\to2$}

\newcommand{\n}{NGC\thinspace~}

\lefthead{Giannakopoulou-Creighton et al.}
\righthead{Star Formation in Giant \hii\ Regions of M101}

\begin{document}

\title{Star Formation in the Giant H\thinspace{\normalsize II}
  Regions of M101}

\author{Jean Giannakopoulou-Creighton\altaffilmark{1,2,3},
        Michel Fich\altaffilmark{1}
        and
        Christine D. Wilson\altaffilmark{4,3}}

\altaffiltext{1}{Department of Physics, University of Waterloo,
  Waterloo, ON N2L 3G1, Canada}
\altaffiltext{2}{Infrared Processing and Analysis Center,
  Jet Propulsion Laboratory,
  California Institute of Technology, Pasadena, CA 91125}  
\altaffiltext{3}{Division of Physics, Mathematics and Astronomy,
  California Institute of Technology, Pasadena, CA 91125}
\altaffiltext{4}{Department of Physics and Astronomy,
  McMaster University, Hamilton, ON L8S 4M1, Canada.}

\begin{abstract}
The molecular components of three giant \hii\ regions (NGC~5461, NGC~5462,
NGC~5471) in the galaxy M101 are investigated with new observations from single
dish telescopes (James Clerk Maxwell Telescope and the NRAO 12-meter) and from 
the Owens Valley millimeter
array.  Of the three \hii\ regions, only NGC~5461 had previously been detected
in CO emission.

We calculate preliminary values for the molecular mass of the GMCs in NGC~5461 by
assuming a CO-to-H$_2$ factor ($X$ factor) and then compare these values with
the virial masses.  We find that the appropriate $X$ factor
is 5 times smaller than the $X$ factor in the Milky Way despite the lower 
metallicity of M101. 
We conclude that the data in this paper demonstrate for the first time
that the value of $X$ may decrease in regions with intense star formation.

 The molecular mass for the \emph{association} of clouds in NGC~5461 is
approximately $3 \times 10^7$ \smass\ and is accompanied by 1--2 times as much atomic mass.
The observed CO emission in NGC~5461 is an order of magnitude stronger 
than in NGC~5462, while it was
not possible to detect molecular gas toward NGC~5471 with the JCMT\@.  An even
larger  ratio of atomic to molecular gas in NGC~5471 was observed, which 
might be attributed to
efficient conversion of molecular to atomic gas.

The masses of the \emph{individual} clouds in NGC~5461, which are
gravitationally bound, cover a range of ($2 \mbox{ -- } 8) \times 10^5$ 
\smass, comparable with the masses of Galactic giant molecular clouds (GMCs).
Higher star forming efficiencies, and not massive clouds, appear to be the prerequisite
for the formation of the large number of stars whose radiation is required to 
produce the giant \hii\ regions in M101.
 
\end{abstract}



\section{Introduction}

Giant \hii\ regions are the most spectacular star-forming regions in normal
galaxies and have been the object of many studies because of their brightness
\cite{1}.  Examples of giant \hii\ regions are found in galaxies of different
brightness and morphology (e.g., Pellet 550 in M31, NGC 604 in M33,  
30 Doradus in the Large Magellanic Cloud, and NGC 5461 in M101).  
Especially impressive are the \hii\ regions observed
in M101, which is a relatively nearby Sc spiral galaxy at $7.4 \pm 0.6$
Mpc \cite{126}.  NGC~5471, one of the giant \hii\ regions in M101, is two orders
of magnitude larger and brighter than W49, the largest
\hii\ region in the Milky Way.  What is special about M101 that it
produces such bright \hii\ regions?  One hypothesis is that these regions result
from the unusual properties of the molecular gas from which the stars that
ionize the gas originate \cite{4}.  In this paper, we present data
on the physical properties of the molecular gas in the giant \HII\ regions of
M101 and discuss the implications for the formation of giant \HII\ regions.

To explain the presence of massive star formation in the clouds associated with
the giant \hii\ regions, Kenney et al.\@  \cite*{10} proposed that either the
initial mass function is enhanced in massive stars, or the gas is consumed more
efficiently in these regions.  The first idea has been investigated by Rosa
and Benvenuti \cite*{41} in
a study of four giant \HII\ regions in M101 with the Faint-Object Spectrometer on
the Hubble Space Telescope.  They concluded that the initial mass function for stars
with masses larger than 2 \smass\ is similar to that in the Solar neighborhood.
This result is consistent with more recent studies of OB associations in inner
part of M101 \cite{112}.  
On the other hand, there was an early indication that the star formation efficiency 
of
massive stars in NGC~5461 is higher than the typical \emph{total} star formation
efficiencies observed in our Galaxy \cite{3}; however, the calculations of the
molecular masses were rather crude.
In \S\ref{s:star}, we find from new observations that the star formation
efficiency is, indeed, \emph{larger} than the efficiencies in
star-forming regions of the Galaxy.  The conclusion in this paper is that higher
star-formation efficiency is the key to the formation of giant \hii\ regions in M101.

We present new observations of three giant \hii\ regions  (NGC~5461, NGC~5471, and
NGC~5462) from the James Clerk Maxwell Telescope\footnote{The JCMT is operated by 
the Joint Astronomy Centre in Hilo, Hawaii
on behalf of the parent organizations Particle Physics and Astronomy Research
Council in the United Kingdom, the National Research Council of Canada,
and The Netherlands Organization for Scientific Research.} (JCMT) in 
\S\ref{s:jcmt}, the NRAO 12-meter Telescope in \S\ref{s:nrao}, and the 
Owens Valley millimeter array in \S\ref{s:ovro}.  These three regions are the
brightest regions in M101 \cite{11}.  The characteristics of the \emph{individual} 
giant molecular clouds (GMCs)  in NGC~5461 are discussed in \S\ref{s:properties} to deduce the molecular
mass of the gas in the \emph{associations} of GMCs.  Both LTE (local thermodynamic 
equilibrium) and LVG (large velocity
gradient) methods, discussed in \S\ref{lte} and \S\ref{lvg},  are used to find the physical
properties of the associations of GMCs. 
 In \S\ref{s:massestemp},
we compare the gas masses and temperatures of the three giant \hii\ regions. 
  We discuss the effect of star formation on the interstellar medium in \S\ref{s:star} 
  and present our conclusions in \S\ref{s:conclusions}.


\section{Observations} \label{s:observations} %

\subsection{JCMT Data and Analysis} \label{s:jcmt}

Receiver A2, used to observe the \COb\ line, has a full-width at half-maximum (FWHM) of $20''$ and an efficiency
$\eta_{fss}$ of 0.80, 
which is a correction for the radiation lost due to forward
scattering and spillover.
Receivers B3i and B3 were both used to observe the \COc\ line; their beam sizes are 
$14''$ and $13''$ respectively, while their $\eta_{fss}$ efficiencies are
0.70 and 0.75.  
All the observations at the JCMT were obtained between 1995 January and 1997 February
in position-switching mode.  The typical single sideband system temperatures
were $300 \mbox{ -- } 400$ K for A2, $800 \mbox{ -- } 1000$ K for B3i, and
$400 \mbox { -- } 600$ K for B3. 
To reduce the data, we used SPECX, which is a spectral line data reduction
code written by Rachael Padman (Cavendish Laboratory, Cambridge, U.K.)  for JCMT
data.  Linear baselines were removed after all the scans of the same position
were averaged.  The data were binned to a frequency resolution of 5 MHz (which
corresponds to 6.50 km s$^{-1}$ at 230 GHz and to 4.35 km s$^{-1}$ at 345 GHz)
to achieve satisfactory noise levels.

\begin{figure}
\figurenum{1}
\figcaption{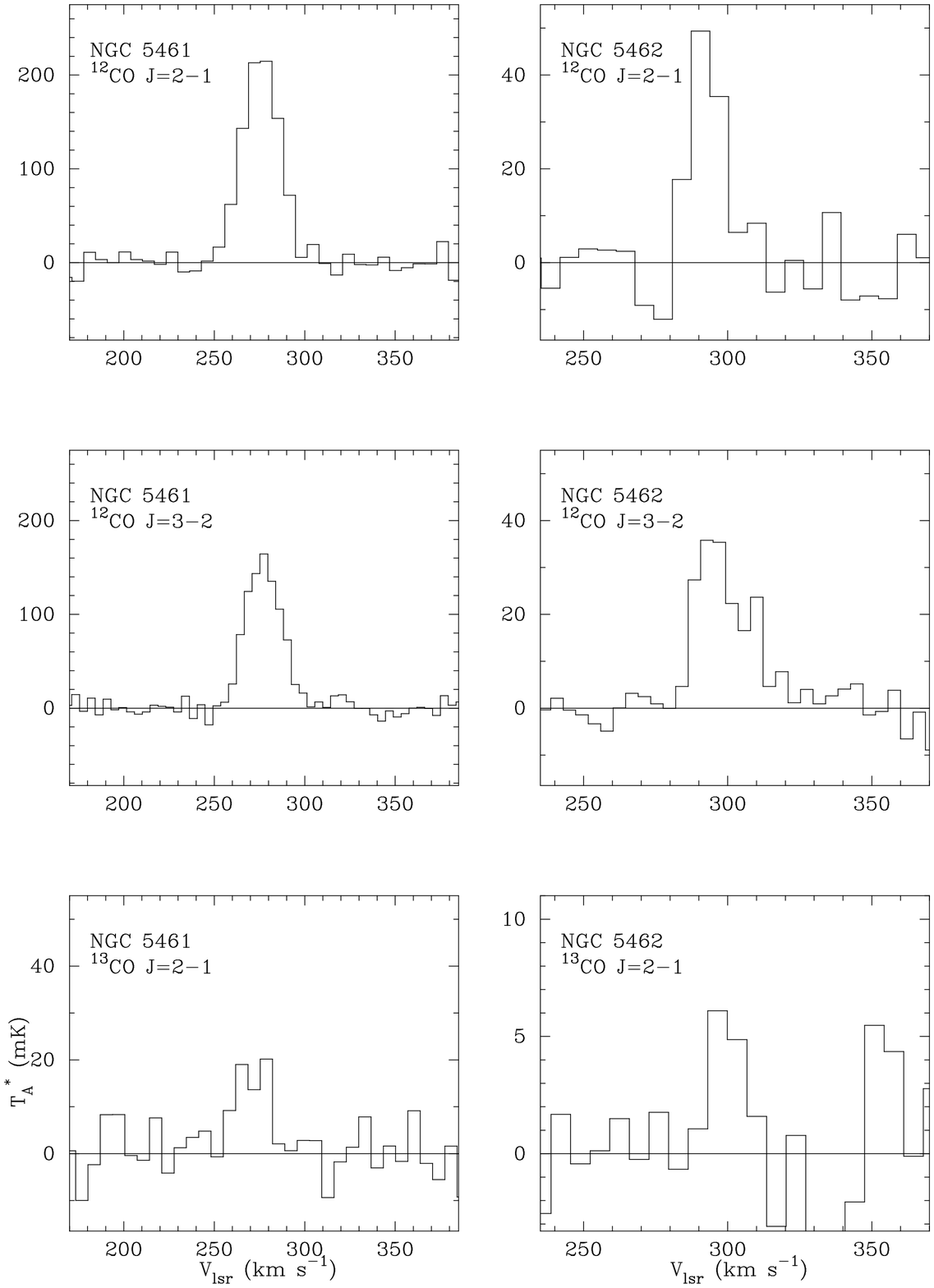}
\caption[JCMT spectra convolved to a $22''$ beam]{JCMT spectra of NGC~5461 and NGC~5462 
convolved  to a $22''$ beam.\label{fig:jcmt}}
\end{figure}

Tables~\ref{tab:NGC5461} and \ref{tab:NGC5462} present the observed integrated
antenna temperature, $\int{T_A^* \; dv}$, for various positions around
 each \hii\ region, while Figure~\ref{fig:jcmt} shows the three spectra
 at the peak positions in NGC~5461 and NGC~5462.   In addition, Tables~\ref{tab:NGC5461} and
\ref{tab:NGC5462} list the parameters obtained from a Gaussian fit to each
of the spectra, where  $V_{peak}$ is the velocity for the maximum
antenna temperature, $T_{peak}$, and 
$\Delta V$ is the velocity width at half-maximum antenna temperature.  These velocity widths are large as 
noted
by Skillman and Balick \cite*{100}.
The uncertainties for each measurement of the line strength have been included
in Tables~\ref{tab:NGC5461} and \ref{tab:NGC5462}; the central velocity is not uncertain to more
than a few percent, while the uncertainty in the velocity width is approximately
 2--3 
km s$^{-1}$.  In the situations where the signal-to-noise ratio is low,
we estimated an upper limit for the integrated intensity by choosing a velocity
interval around the central velocity that maximizes the value of the integrated
intensity.

\placefigure{fig:jcmt}

The uncertainties for the peak temperature, $T_{peak}$,
quoted in Tables~\ref{tab:NGC5461} and \ref{tab:NGC5462} are calculated
for the smoothed data.  
The uncertainties range between
$7 \mbox{ -- } 24$ mK for $^{12}$CO and $ 2.5 \mbox{ -- }4 $ mK for $^{13}$CO\@.
In addition to the random uncertainties, it is possible that there are
systematic uncertainties.  To check for systematic differences among observing
sessions, we compared the data from different runs for positions where there are
spectra with a high signal-to-noise ratio, for example, a calibrator or
NGC~5461.  The variations in flux among
observing sessions are typically 10\%, so there could be a systematic
error of 10\% that is introduced from the flux calibration.

\placetable{tab:NGC5461}

\placetable{tab:NGC5462}

$^{12}$CO $J=2\to1$ emission was detected at the $7 \, \sigma$ level or
better toward all eleven points around NGC~5461.  The peak antenna temperatures
ranged between 68 and 228 mK\@.
Interestingly, the $^{12}$CO $J=3\to2$ (345 GHz) emission from three ($a,b,c$)
out of the five central points observed in NGC~5461 was between $65 \mbox{ -- }
100$\% as strong as the emission seen in the $^{12}$CO $J=2\to1$ spectra
(Table~\ref{tab:NGC5461}). 

 The strongest detections in NGC~5462 in both \COb\ and $J=3\to2$ were for
position $g$ and not for $a$, which coincides with the center of the region.
Special care was taken so that four positions around NGC~5462g were observed 
in \COc\ in
order to be convolved together so that they could be compared with the lower
resolution \COb\ data.
Finally, it was possible to get a significant detection of \tCOb\ toward NGC~5461a
(Table~\ref{tab:NGC5461}), while the detection
toward NGC~5462g is more questionable (Table~\ref{tab:NGC5462}). 
 
 To our surprise, there was no detection of $^{12}$CO $J=2\to1$ toward 
 NGC~5471, one of the brightest
\hii\ regions in M101.  To calculate an upper limit for the integrated intensity
from the five positions observed, we added the positive 
integrated intensity only.  The upper
limit derived in this way for NGC~5471 is 0.70 K km s$^{-1}$ integrated over 40 km
s$^{-1}$.

\subsection{National Radio Observatory (NRAO) 12-meter}
\label{s:nrao}

On 1996 November 6, we used the NRAO 12-meter telescope to observe NGC~5461,
NGC~5462, and NGC~5471 in the rotational transition of $^{12}$CO $J=1\to0$
(Table~\ref{tab:nrao}).  The beam is large enough ($55''$) to encompass each of
the three giant \HII\ regions; the integration times were 60 minutes, 120
minutes, and 78 minutes for NGC~5461, NGC~5462, and NGC~5471, respectively.
Each individual scan was 6 minutes.  Typical system temperatures at 115 GHz were
$350\mbox{ -- }400$ K\@.  The 256 channel 1 MHz (2.6 km s$^{-1}$) dual
polarization filterbank was configured in series mode to gain a factor of $\sqrt{2}$ in
the noise level.  The telescope software presents the data in units of corrected 
radiation temperature, $T_R^*$, as
opposed to the corrected antenna temperature, $T_A^*$, that the JCMT system delivers.
The telescope scale was checked by observing K3-50 (observed peak $T_R^\ast= 24
\pm 0.1$ K); the uncertainty in the $T_R^\ast$ calibration was estimated to
be about 15\%.

\placetable{tab:nrao}

\begin{figure}
\figurenum{2}
\figcaption{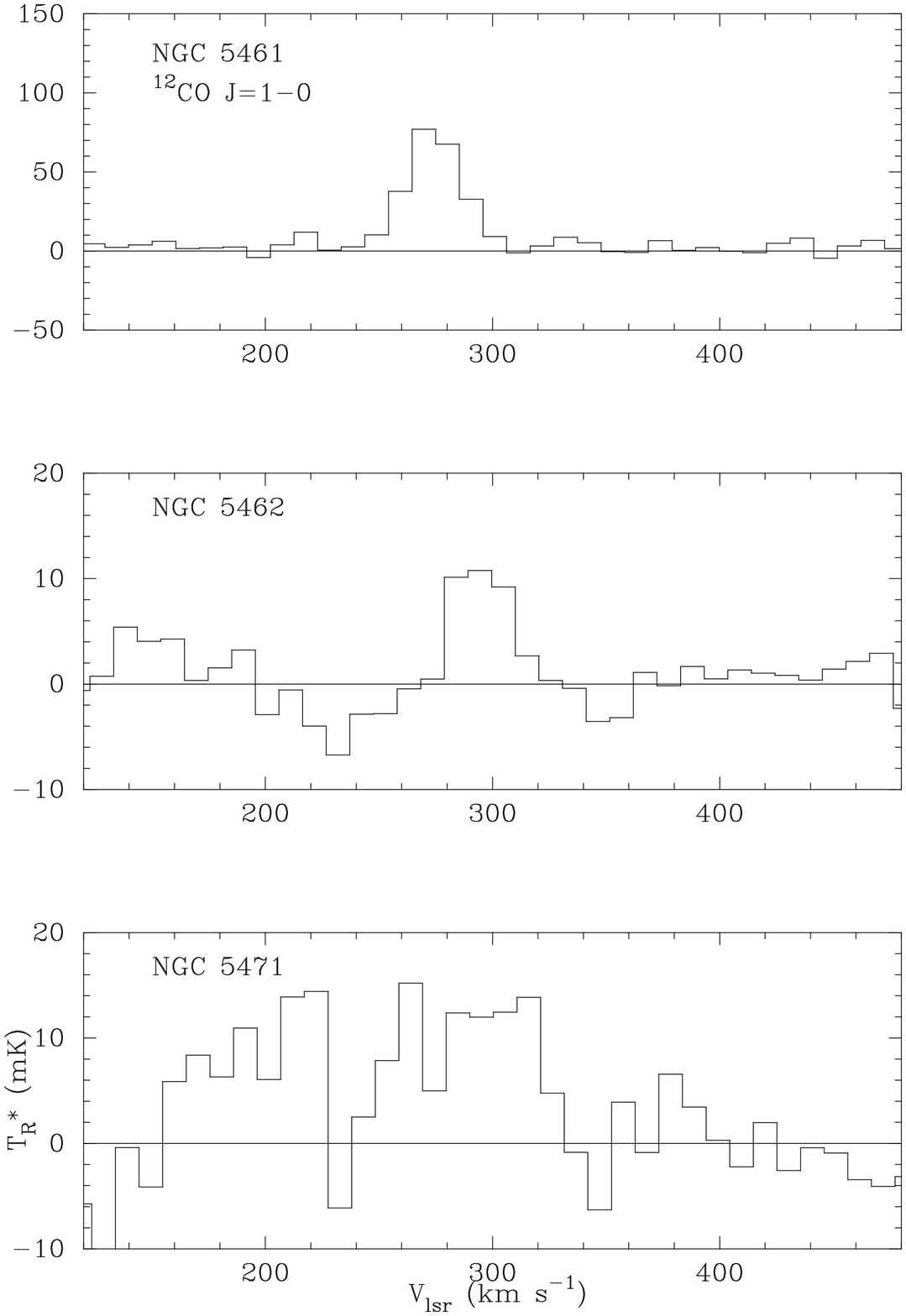}
\caption[NRAO spectra for the three giant \hii\ regions]{NRAO spectra
of \COa\ toward NGC~5461, NGC~5462, and NGC~5471.  
\label{fig:nrao}}
\end{figure}

The data were reduced with UniPOPS\@.   Linear baselines were removed from two of the spectra
(\n5461 and \n5471) while a polynomial baseline was removed from
the spectrum of \n5462.  The data were smoothed 
to a resolution of 10 km s$^{-1}$ (or 4 MHz at 115 GHz) to improve
the signal-to-noise ratio.  The \COa\ spectra are shown in Figure~\ref{fig:nrao},
and the integrated intensities are given in Table~\ref{tab:nrao}. 

\placefigure{fig:nrao}  

\subsection{Owens Valley Millimeter-Wave Interferometer Data}
\label{s:ovro}

NGC~5461 and NGC~5462 were observed with the Owens Valley (OVRO) Millimeter-Wave
Interferometer during 1996 February and April and during 1997 February and April.  This millimeter
array has 6 antennae, which have diameters of 10.4 m.  
Two configurations of the array (A and C) were combined so that the synthesized
beam is $2\farcs62 \times 2\farcs03$ for NGC~5461 and $3\farcs14 \times
2\farcs73$ for NGC~5462. The total track length was 16 hours for
each region in each configuration.
Typical single-sideband system temperatures at the zenith were $600 \mbox{ -- }
1000$ K\@.  All four independent correlator modules of the digital spectrometer
system were used to observe the \COa\ line with an effective bandwidth of 126
MHz (328 km s$^{-1}$) and a resolution of 1 MHz 
(2.6 km s$^{-1}$).

To reduce the data from the OVRO interferometer, we used the \texttt{mma}
software package, which is written and maintained by the Caltech millimeter
interferometry group \cite{173}.  For the two tracks of NGC~5461, Neptune and the quasar
3C273 were used for the flux calibration measurements, which agreed to within $20\%$.
For the two tracks of NGC~5462,  one was calibrated with Uranus observations, and
the other with the quasar 3C345.  The flux measurements from these two tracks agreed
to within $10\%$.

The gain calibrator was the quasar 1418+546
[$\alpha(1950)=14\timehr18\timemin06\fs200$ and
$\delta(1950)=+54\arcdeg36\arcmin57\farcs80$], which is unfortunately fairly
weak.  The average measured flux during the NGC~5461
observations was 0.80 Jy, while the flux during the NGC~5462
observations a year later was 0.55 Jy.  Since it is possible that the
intrinsic brightness of the gain calibrator changed during the course
of a year, we adopt these two separate values for the remaining
analysis.

The data sets were edited to remove poor data with the main criterion 
being that the coherence on the gain calibrator was higher than 50\%.  We determined 
passband calibration for the NGC~5461 tracks using the quasars 3C273 and
3C454.3.  For the NGC~5462 tracks, the quasars 3C454.3 and 3C345 were
used for passband calibration.

After the basic reduction was completed, we used Miriad \cite{178} to map and clean the data.
Because the signal-to-noise ratio was relatively small, we used natural weighting.
All channel maps were cleaned to the $1.5 \, \sigma$ level with fewer than 1000
iterations.  The rms noise and the maximum signal for the maps integrated over 52 km
s$^{-1}$ were 0.025 Jy beam$^{-1}$ and 0.12 Jy beam$^{-1}$ for NGC~5461, and 0.022
Jy beam$^{-1}$ and 0.09 Jy beam$^{-1}$ for NGC~5462.  The rms noise and the maximum
signal for the 1 MHz channel maps were 0.10 Jy beam$^{-1}$ and 0.36 Jy beam$^{-1}$
for NGC~5461, and  0.10 Jy beam$^{-1}$ and 0.33 Jy beam$^{-1}$ for NGC~5462.

Once the cleaning process was completed, we identified candidate GMCs in the maps.
Three different types of plots were used.  The \emph{integrated
map} is a plot that has been integrated over a wide velocity range (52 km s$^{-1}$)
to include all the emission.  The \emph{channel maps} are a series of maps where the
region is plotted by integrating over one channel only (2.6 km s$^{-1}$).  Finally,
the \emph{optimum map} for each cloud is integrated only over the velocity range
in which a given feature is visible at the $3 \, \sigma$ level or better.

It must be stated in no uncertain terms that the process of identifying GMCs is
fairly subjective.  We have chosen three relatively conservative criteria so that
the results inspire some confidence.  The first criterion used to identify the GMCs
is that the feature should be at least $2 \, \sigma$ in the integrated map.  The
second criterion required features to persist at the $3 \, \sigma$ level over
two consecutive channels; this
method, however, could result in underestimating the number of clouds, especially
the ones with a narrow velocity width.  The best velocity range for each GMC
candidate was determined from the channel maps, and for each feature, the optimum
map integrated for the appropriate velocity range was generated.  The third
criterion required the feature to have signal of at least $3 \, \sigma$ in the
optimum map.  It is possible for the second criterion to be satisfied but not the
third if the features drift slightly from channel-to-channel so that the `optimum'
integrated flux is less than 3$ \, \sigma$ above the noise.

\begin{figure}
\figurenum{3}
\figcaption{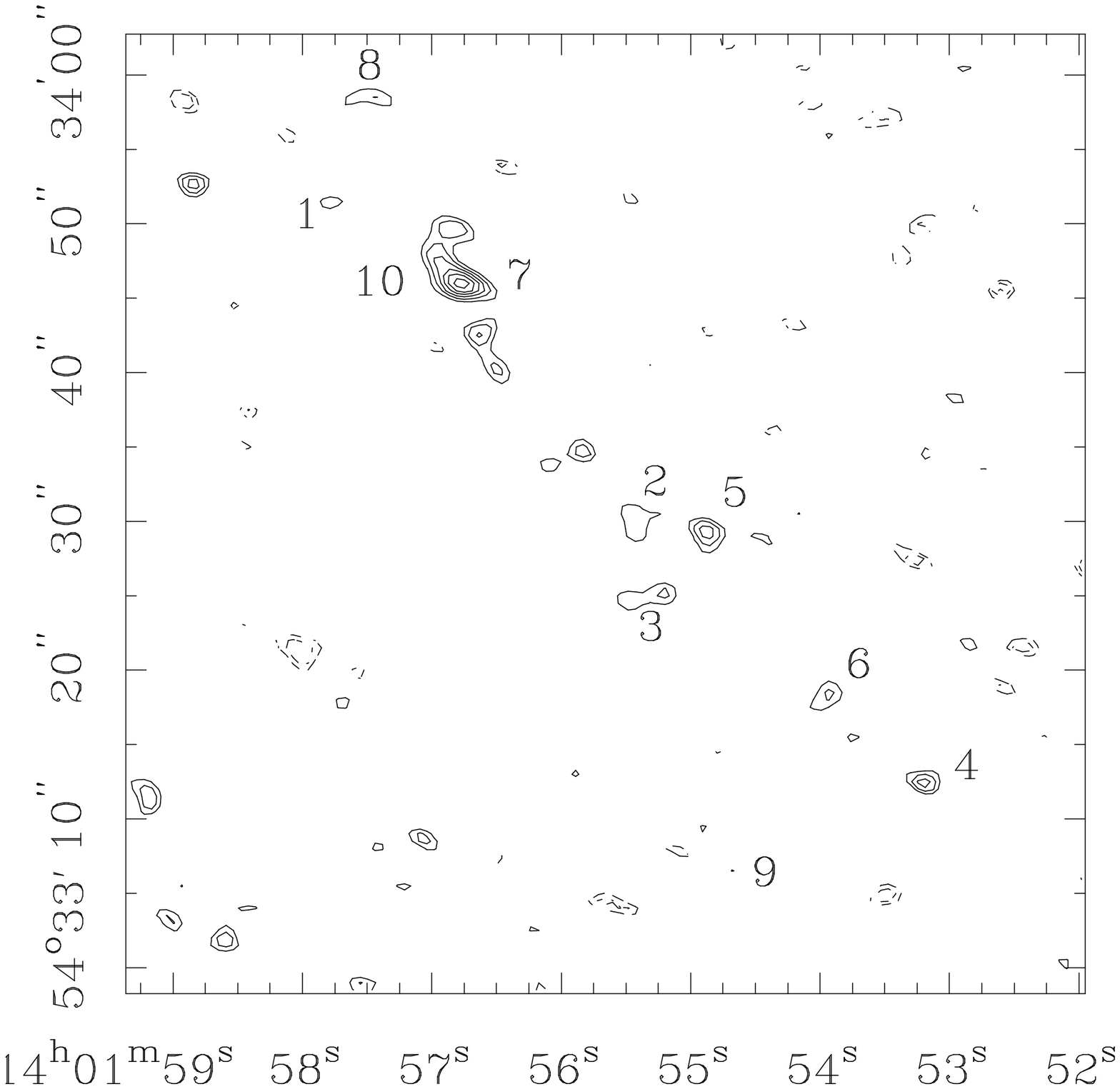}
\caption[Integrated OVRO map of NGC~5461 over 52 km
s$^{-1}$]{\COa\ map of NGC~5461 obtained with the OVRO Millimeter-Wave
Array integrated over 52 km s$^{-1}$.
The contours have been plotted at the $\pm 2 \, \sigma$, $\pm 2.5 \,
\sigma$, $\pm 3 \, \sigma$, $ \pm 3.5 \, \sigma$, $ \pm 4 \, \sigma$,
and $ \pm 4.5 \, \sigma$; positive contours are solid while negative
contours are dotted.  The rms noise is 0.024 Jy beam$^{-1}$.  The number 
beside each peak identifies individual
GMCs; clouds 7 and 10 coincide spatially, but they have
different velocities (Table~\ref{tab:ovro_gma}). The fluxes
in this figure have not been corrected for the falloff of the
primary beam.\label{fig:61int}}
\end{figure}

The integrated map of NGC~5461 (Figure~\ref{fig:61int}) has thirty-three
features with peak fluxes of at least $2 \, \sigma$.  Of the thirty-three features
that appear in the integrated map, nine can be seen at the $3 \, \sigma$ level
in two consecutive channels of the channel maps.  In addition, we have also
considered two other features (4 and 6 in Figure~\ref{fig:61int}) that appear on
the integrated map and in three consecutive channels at the $2 \, \sigma$ level.
Of the eleven GMC candidates, one is eliminated by the third criterion, i.e.,
the total flux in the optimum map is not three times higher than the noise of
the optimum map.  The rms noise determined in the optimum maps varies between
$0.05 \mbox{ -- } 0.069$ Jy beam$^{-1}$ depending on the number of channels over
which the signal has been integrated.

\placefigure{fig:61int}

The characteristics of the ten GMCs are given in Table~\ref{tab:ovro_gma}.  The
positions were determined from the optimum maps.  The sizes are not deconvolved
from the synthesized beam; no cloud appears significantly larger than
the beam in both dimensions, and thus we can only place upper limits on the
true size of the clouds.  The
cloud positions are given for the center of the peak of the GMC in terms of offsets
from the field center with an estimated uncertainty of $0\farcs5$. Two features, 
clouds~7 and 10, have similar coordinates, but they are separated in velocity 
space.   Table~\ref{tab:ovro_gma} also includes the integrated flux
measured from the optimum map.  We calculated the equivalent brightness temperature,
$T_B$, by multiplying the peak flux in a one channel map by the conversion factor
from Janskys to Kelvin (17.34 K/Jy).  These brightness temperatures 
(4.5 -- 8.1 K) are larger than those found in the GMCs toward
M33 \cite{34} despite the fact that the GMCs in M33 are resolved and much
closer.  These large brightness temperatures suggest that the GMCs in M101 
are close to being resolved.

\placetable{tab:ovro_gma}

The emission from NGC~5462 is much weaker than that from NGC~5461; in fact, the
emission is so much weaker that there are no features that appear in consecutive
channels at the $3 \, \sigma$ level.  If all the positive features are added, then
the upper limit to the total flux from the region is  $<16$ Jy km s$^{-1}$.

\section{Molecular mass from the Empirical Method} \label{s:properties}

\subsection{Mass of individual GMCs in NGC~5461}
\label{empirical}

The column density can be calculated via an empirical relation based on data
that suggest that the column density of hydrogen, $N_{H_2}$, is linearly proportional
to the observed radiation temperature integrated over the emission line of
$^{12}$CO $J=1\to0$ \cite{71,139}.  The constant of proportionality, $X$, 
is defined by \cite{153}
\begin{equation}
	X = {N_{H_2} \over \int T_R^* \, dv }.
	\label{a_twentyfive}
\end{equation}
The range of $X_{Gal}$ found in the literature is 1 -- 12
$\times 10^{20}$ cm$^{-2} \,$ (K km s$^{-1}$)$^{-1}$   \cite{139}, and  
we assume a value of $X_{Gal}= 3 \times 10^{20}$
cm$^{-2} \,$ (K km s$^{-1}$)$^{-1}$. However, the $X$ factor probably depends
upon the metallicity of the gas. 
It has been suggested that galaxies with lower metallicity than that of the 
Milky Way have higher values of
$X$  \cite{37} with
\begin{eqnarray}
\log(X/X_{Gal})&=&(5.95 \pm 0.86)\nonumber\\
  &&\quad-(0.67 \pm 0.10) [12+ \log({\rm O}/{\rm H})].
\label{a_twentyfivea}
\end{eqnarray}
The giant \hii\ regions of M101 have low metallicities, so the value
of $X$ should be adjusted: the value of [12+ log(O/H)] for NGC~5461 
is $8.39 \pm 0.08$, while for NGC~5471 it is $8.05 \pm 0.05$ \cite{14}.

The uncertainties introduced by the value of $X$ may be severe, especially in
a galaxy with intense star formation \cite{159}.  However, the use of the $X$
factor is a standard way to estimate molecular cloud masses, so we adopt
 an initial
value of $X$ correcting for the known metallicity of these regions and compare the resulting 
molecular gas mass with another, usually more reliable, virial mass to determine an 
appropriate value of $X$ for the \hii\ regions in M101.  
As a starting point, we use the value 
of $X_{NGC~5461}=6 \times 10^{20}$ cm$^{-2} \,$ (K km
s$^{-1}$)$^{-1}$ for NGC~5461.   The hydrogen
column
density can be obtained from equation (\ref{a_twentyfive}), and the
molecular mass in terms of flux density, $S_{\nu}$, can be written as \cite{34}
\begin{eqnarray}
	M_{mol} =&& 1.61 \times 10^4 \, \smass\
\left(    
{X \over {3 \times  10^{20} {\rm cm}^{-2} \, (\rm{K} \, \rm{km} \, 
\rm{s}^{-1})^{-1}}}
\right) \nonumber \\
&&\quad\times 
\left({d \over {\rm Mpc}}\right)^2 
 \int {S_{\nu} \over {\rm Jy}} \, {dv \over {\rm km} \,  {\rm s}^{-1}}.
\label{a_twentysevena}
\end{eqnarray}
We use 
equation (\ref{a_twentysevena}) to calculate the masses of the individual
GMCs (Table~\ref{tab:sizeline}); we have corrected the fluxes (and masses)
for the primary beam falloff.  With this equation, the individual features in
NGC~5461 have masses of $(2\mbox{ -- }9) \times 10^6$ \smass\ and appear to be much
more massive than the GMCs in the Milky Way, which have
masses typically around $10^5$ \smass.   

\placetable{tab:sizeline}

As a check on these masses, we compare the molecular masses to
the virial masses \cite{179}. If the virial mass is less than or equal to
the molecular mass,
then the clouds are considered gravitationally bound.
The virial mass is adapted [by converting $\sigma_{1D}$ to
$\Delta V_{\rm FWHM}$] from Rand \cite*{33}:
  \begin{equation}
M_{vir} = 95 \, \smass\ 
\left({\Delta V_{{\rm FWHM}} \over {\rm km \, s^{-1}}}\right)^2
\left({D \over {\rm pc}}\right).
\label{d_two}
\end{equation}
We define the diameter, $D$, to be $1.4 \, D_{FWHM}$ to include all the emission 
from the GMC and for consistency with previous studies \cite{34}.  
In our calculations, we have assumed an upper limit for $D_{FWHM}$ of 80 pc
because none of the 
GMCs are resolved from the beam in both dimensions;
therefore, the virial masses estimated are upper limits based on a
diameter of 110 pc. 
 These masses have been tabulated in
Table~\ref{tab:sizeline} along with the masses found by the empirical
method and the expected diameter, $D_{exp}$, given the observed velocity 
dispersion, $\Delta V_{obs}$ and the size:line-width relation \cite{118,160}: 
   \begin{equation}
{\Delta V \over {\rm km \, s^{-1}}}= 1.2 \biggl({D \over {\rm pc}}\biggr)^{0.5}.
\label{d_one}
\end{equation}

 We find that the virial mass of each cloud
is generally an order of magnitude smaller than the molecular mass calculated
from the metallicity corrected $X$ factor and is
similar to the masses of the GMCs in the Milky Way.  We have more confidence
in the universality of the virial theorem and, thus,
this large discrepancy suggests that a smaller value of $X$ for the 
giant \hii\ regions in M101 is more appropriate.
The physical justification for adopting a different value of $X$ is that
$X$ depends on temperature, and the observed high brightness temperature indicates
that the gas is
hot.  If these clouds obey the Galactic size:line-width relation, they
have rather low filling factors within the OVRO beam and their true
brightness temperatures would be even higher than what is observed. It is possible
 that the high temperature of the
gas lowers the value of $X$ in a star forming region \cite{159}.
The average value of the ratio $M_{mol}$/$M_{vir}$ for the GMCs in NGC~5461
is $10 \pm 4$,  which suggests
that the appropriate value of $X_{NGC~5461}$  is 
$6  \times 10^{19}$ cm$^{-2} \,$ (K km s$^{-1}$)$^{-1}$ or approximately
five times smaller than the canonical Galactic value (and ten times smaller
than the metallicity corrected value).  We adopt the virial mass as the most
correct estimate of the mass of the GMCs.  These observations
are the first to demonstrate a clear decrease in the value of $X$ due
to heating by intense star formation.  We adopt the same value for $X_{NGC~5462}$ because the two giant \hii\ regions have similar
metallicities.  \label{Xranges} For NGC~5471, we use the value $X_{NGC~5471}$ of
$12 \times 10^{19}$ cm$^{-2} \,$ (K km s$^{-1}$)$^{-1}$ because of its lower
metallicity.

\subsection{Mass of the Associations of GMCs}

After examining the masses of the individual GMCs, we use equation
(\ref{a_twentysevena}) to calculate the mass of the entire association
of GMCs for each region using the appropriate value of $X$.
With observations from the NRAO 12-meter telescope (Table~\ref{tab:nrao}), we
obtain the integrated intensity by multiplying the integrated 
$T_R^*$ with the factor 34 Jy/K\@.  From the integrated intensity, we
calculate the total molecular mass from equation~(\ref{a_twentysevena})
(Table~\ref{tab:ansummary}). 

\placetable{tab:ansummary}

The integrated flux from the OVRO map was measured from the integrated 
map over a velocity range of 52 km
s$^{-1}$ of \COa\ emission.  The value for the integrated flux density is $33
\pm 13$ Jy km s$^{-1}$, which corresponds to $(60 \pm 23) \times 10^5$
\smass.
 This result from the OVRO millimeter array is smaller than the mass 
obtained from the single dish data (12-meter
telescope).  To explain the difference, one might evoke
the presence of a constant, broadly distributed contribution to the intensity,
which would have been undetected by the interferometer \cite{34}. The same
explanation might be appropriate for the measurements of NGC~5462 
(Table~\ref{tab:ansummary}).

 The mass of molecular gas associated with NGC~5461 is
greater than the masses associated with the other two \hii\ regions.  Although the
calculations indicate that NGC~5462 has a slightly smaller molecular mass than
NGC~5471, their allowed values are the same to within their uncertainties.  The
uncertainties in the total molecular mass are derived from the uncertainty in the
integrated radiation temperature; the (potentially more significant) systematic
uncertainty in the value of $X$ has been ignored.

\section{LTE Analysis}  \label{lte}%

In this section, we calculate  the optical depth and the column
density based on the assumption that the gas is in 
\emph{local thermodynamic equilibrium}
(LTE)\@.  For the LTE analysis, we used $^{12}$CO and $^{13}$CO $J=2\to1$ data
obtained at the JCMT\@. 
The LTE method is frequently used to calculate the physical properties of
molecular gas especially when isotopomers of CO or of other molecules such as CS
and NH$_3$ have been observed [e.g.,  Lada and Fich \cite*{161}; 
Giannakopoulou et al.\@  \cite*{162}].
 The $^{12}$CO $J=2 \to 1$ and $^{13}$CO $J=2 \to 1$ transitions
have very similar frequencies, so the ratio of the transparency of the gas
in the two transitions is approximately given by
\begin{equation}
{1-e^{- {}^{12}\tau_{\nu}} \over   1-e^{- {}^{13}\tau_{\nu}}} =
{  k \, ^{12} T_R / h \,\nu_{12} -e^{-h \, \nu_{12} / k \, T_{bg}}  \over 
   k \, ^{13} T_R / h\,  \nu_{13} -e^{-h \, \nu_{13} / k \, T_{bg}}    }, 
\label{a_nine}
\end{equation}
where $\nu$ is the frequency of the transition, $\tau_{\nu}$ is the optical depth, 
$T_R$ is the radiation temperature, and  $T_{bg}=2.73$ K is the temperature of the
microwave background.

We can calculate $^{13}\tau_{\nu}$ if we assume a value for the abundance ratio 
$\psi = [^{12}$CO]/[$^{13}$CO].  This ratio may increase with the galactocentric 
distance \cite{79}:  $\psi$ has been measured to be as low as 24
within 4 kpc and as high as 79 at 12 kpc 
from the center of our Galaxy. For the distance of the two giant \HII\ regions 
from the center of M101 (between 4 and 6 kpc), we expect
the most appropriate ratio to be at the lower end of these values.  \label{psivalues}

Since we have assumed that the gas is in LTE, the kinetic temperature $T_k$ is
equal to the excitation temperature and is given by
\begin{equation}
T_k = \frac{h\nu/k}{\ln\{ (1-e^{-\tau}) [kT_R/h\nu -
 (e^{h\nu/kT_{bg}}-1)^{-1}]^{-1} +1\}}.  
	\label{a_eleven} 
\end{equation} 
Appendix A contains a detailed
discussion of how we obtained the radiation temperature.
The values of the kinetic temperature are between 8 and 25 K, which
are fairly low for a star forming region.  Even lower values of 4--6 K are
found for NGC~5462.  However, these are average temperatures
over large volumes of gas. Near young stars, the temperature of the gas 
will likely be much higher.  The smaller value of $X$ found for the GMCs
in NGC~5461 is possible evidence of higher temperatures, as are the
large brightness temperatures of 4 -- 8 K observed in the \COa\ line.

From references such as Mitchell et al.\@  \cite*{138} and Giannakopoulou
et al.\@ \cite*{162}, we find that the total
integrated $^{13}$CO column density, $N_{^{13}CO}$, is given by 
\begin{eqnarray}
	N_{^{13}CO} =&&
{3k \over 8  B_r  \pi^3  \mu^2}  \int  
{e^{h  B_r  l(l+1)/k  T_k} \over (l+1)} \nonumber \\
&&\qquad\times
{T_k + {h  B_r / 3 k} \over 1-e^{-h  \nu /k  T_k}}\, {}^{13}\tau_v  dv,
	\label{a_twentyone}
\end{eqnarray}
where the rotational moment of CO is $B_r=5.764 \times 10^{10}$ Hz and the
electric dipole moment of CO is $\mu=1.12 \times 10^{-19}$ esu cm \cite{165}.
For the error analysis of the column density, we use a robust bootstrap
method \cite{168}.  \label{bootstrap} We estimated the probability distribution
of the column density, via equation~(\ref{a_twentyone}), by a Monte Carlo
method in which Gaussian random numbers representing  $^{12}T_A^*$ and $^{13}T_A^*$ 
are generated with population means equal to the measured temperatures
and with population standard deviations equal to the measured uncertainties in
the temperatures.   From the characteristics of these distributions, one can
obtain the uncertainty that is associated with the total column density.

To obtain the column density $N_{^{13}CO}$ for each region, we summed the column
densities per channel over all channels.  For a channel to be included, 
we required that its peak $^{12}$CO
signal is at least $3\sigma$, and its peak $^{13}$CO is at least $2\sigma$. The
median column density for NGC~5461 is $6.5 \times 10^{15}$ cm$^{-2}$.
These criteria result in a lower limit of $1.2 \times 10^{15}$ cm$^{-2}$
 for the value of the column density  (and
consequently of the mass) of NGC~5462, which only has one channel that meets the
restrictions.

The column density is used to obtain the total molecular mass for the three
regions.  Since the main constituent of the interstellar medium is molecular
hydrogen, the $^{13}$CO column density obtained above needs to be extrapolated
to that of $^{12}$CO, $ N_{^{12}CO}$, which in turn will be converted to the
H$_2$ column density, $N_{H_2}$.  The values of the isotopomer ratio, $\psi$, and
of the ratio of the column density of molecular hydrogen to the column density
of CO are uncertain.  We adopt a value of $\psi=20$ because the giant \hii\
regions are fairly near the center of M101 (also \S\ref{lvg}).  In addition, we need to determine
a suitable value for the proportionality constant [$N_{H_2}$]/[$N_{^{12}CO}$].
The canonical value of this parameter is $10^4$ for the Galaxy \cite{152}.
However, this ratio depends on metallicity.  Since NGC~5461 and NGC~5462 are
 4 times more
metal poor, and NGC~5471 is 9 times more metal poor than the Milky Way \cite{14},
we adopt values of $4 \times 10^4$ and $10^5$, respectively, for 
these \hii\ regions.

The mass of the H$_2$ gas \label{hydrogenmass}is found by multiplying the total
H$_2$ column density, $N_{H_2}$, by the beam area, $A$, and the mass of a hydrogen molecule
$m_{H_2}$.  Finally, the fractional helium abundance (10\% by number) is taken
into account in order to obtain the total mass within a beam.  Thus, the
total molecular gas mass for NGC~5461 and NGC~5462 is
\begin{equation} 
	M _{mol}= 6.3 \times 10^{-9} \, \smass \,
  \left( {N_{^{13}CO} \over {\rm cm^{-2}}}\right)
  \left( {A \over {\rm pc^2}} \right).
	\label{a_twentyfour}
\end{equation}
The masses of the molecular gas are given in Table~\ref{tab:ansummary}.
These masses are much larger than Galactic GMCs, which have typical
values of $10^5$ \smass\ \cite{56}.  Large masses are expected from regions as
big as those observed (700 pc or $20''$), which would contain an association of GMCs.

\section[LVG analysis]{LVG analysis}   \label{lvg}

There are two extreme non-LTE approaches in dealing with the radiative transfer
problem.  One approach is to assume that there is no global motion: all the
emission is a result of small-scale thermal motions and turbulence.  The second
approach is to assume that the gas cloud has a large velocity gradient (LVG), which
means that the CO emission from one part of the cloud will be Doppler-shifted to
a frequency that will not be re-absorbed by gas in other parts of the cloud
\cite{60}.  Under these conditions, the emission from a molecule will only be
absorbed by neighboring molecules, and the radiative transfer can be solved
locally \cite{58}.  The first approach is called the microturbulence method, and
the second approach is called the LVG method.
The two non-LTE approaches typically produce results that agree within the
errors of the methods, which are a factor of three due to uncertainties in
geometry \cite{59}. We have used just one non-LTE method, the LVG method, to compare with
our LTE analysis \cite{59,23}.

To estimate the physical conditions in the GMCs of M101, we have
used the LVG code written by Jessica Arlett and Lorne Avery for a spherical,
uniform cloud.  This program computes the radiation temperatures of various
CO transitions for ranges of kinetic temperature, density of H$_2$, and an abundance
parameter, which is defined as $(N_{CO}/N_{H_2})/(dV/dr)$.
These radiation temperatures were fitted to the observed values (Table~\ref{tab:ratios})
by minimizing a $\chi^2$ statistic to give the best fit for kinetic temperature,
 density, and abundance. 
In addition to these parameters, the LVG results depend on the isotopomer ratio, 
$\psi = $[$^{12}$CO]/[$^{13}$CO].   We found that the best fits occur for $\psi=20$,   which
is in agreement with the observed value in the Milky Way \cite{79}.  

\placetable{tab:ratios}

For NGC~5461, we found reasonable fits both for low temperatures ($T_k=30 \mbox{ -- }90$ K)
and for high temperatures ($T_k=230 \mbox{ -- }250$ K); the best fit was for $T_K=60$ K.
However, the best fit for density, $n_{H_2} = 3 \times 10^3$ cm$^{-3}$, was the same
for both ranges of temperature.  A GMC with diameter 112 pc and this density has a mass
of $1.3 \times 10^8$ \smass, which is two orders of magnitude larger than those
obtained with the empirical method (Table~\ref{tab:ansummary}).  The clouds are not resolved, and we do not expect the cloud to be 
completely uniform: the volume-averaged  density
is usually substantially smaller than $10^3$ cm$^{-3}$ \cite{34}.  
By comparing the results from the LVG analysis and the empirical method,
we conclude that the filling factor is on the order of  
 1\% for the clumpy material.  This low filling-factor indicates that 
the gas emitting the CO has formed dense clumps, which are surrounded by a
lower density envelope.  A similar situation is observed in the molecular
gas of 30 Doradus \cite{179}: the intense radiation field is considered the reason for
the dissociation or ionization of the gas in the envelope.  


The LVG analysis cannot be done for NGC~5462 because there are only two
available ratios (Table~\ref{tab:ratios}).  The LVG analysis is not
conclusive with only two ratios \cite{54}, but one can use the available ratios
to make some general statements about the physical properties of the gas.
Because the $^{12}$CO ($J=3 \to 2$)/($J=2 \to 1$) ratio for NGC~5462 is larger than 
that of NGC~5461, the
molecular gas in NGC~5462 may be warmer than the molecular gas around NGC~5461.
\label{a:tempratio} One expects from the Boltzmann equation that, for usual
temperatures of quiescent molecular gas (10 K), the second level is more
populated than the third level, which means that for cold gas the ratio
$^{12}$CO ($J=3 \to 2$)/($J=2 \to 1$) is less than unity; however, in the case of NGC~5462, 
the ratio is
larger than unity, which implies that the gas is probably considerably hotter
than typical quiescent gas.

\section{Masses and Temperatures}  \label{s:massestemp}

NGC~5461 is the brightest
\hii\ region in CO emission; therefore, the data set for NGC~5461 is
the most complete of all the giant \hii\ regions in M101. There are
two estimates (from the LTE and from the empirical estimate using the
NRAO data) for the mass of the association of GMCs in
NGC~5461, which range in value from ($15 \mbox{ -- } 40) \times 10^6$
\smass.  Although the discrepancy of a factor of 3 is not surprising when
we consider the uncertainties in the methods involved, the difference could
be partially due to an underestimate of $\eta_c$ (see Appendix).  It is likely that
there are GMCs in NGC~5461 that were not detected and would have contributed
in increasing the coupling efficiency.

Both the LTE method and the empirical method suggest that NGC~5462 has
roughly an order of magnitude less molecular mass than NGC~5461.   This conclusion is 
consistent with the fact
that the molecular gas in NGC~5462 was not detected with the Owens
Valley array.
NGC~5462 might not be as massive as NGC~5461, but it may be hotter
than NGC~5461 since the $^{12}$CO ($J=3 \to 2$)/($J=2 \to 1$) ratio in NGC~5462 is even higher than
the extraordinary cloud NGC~604-2 in M33. Unfortunately, the uncertainty of
this ratio for NGC~5462 is fairly high; within its uncertainty, it agrees
with normal clouds in M33 and IC10 \cite{156}.

NGC~5471 \label{s:def71} is significantly weaker in CO emission than NGC~5462 and
NGC~5461, so the only observations that were made of NGC~5471 were in \COa\
and $J=2 \to1$ emission.   Unlike NGC~5462, NGC~5471 is several kpc away from
the center of M101 and, therefore, has
considerably lower metallicity than regions closer to the center of
M101.  With the higher adopted value for $X$ for this region,
the upper limit to the molecular mass in NGC~5471 is comparable
to that of NGC~5462. 
 If the $X$ factor is in fact even larger, then the mass of the
molecular gas in NGC~5471 could even be comparable to that of NGC~5461.
However, a more probable explanation for  the deficiency of CO emission in
NGC~5471 is that there is genuinely less molecular gas present.  A
clue to this problem is the presence of large amounts of atomic gas
near the giant \hii\ region \cite{kamphuis:93}.  Perhaps the atomic
gas was converted to molecular gas more slowly in NGC~5471  than in the
other two giant \hii\ regions, or perhaps more of the molecular gas has been
rapidly dissociated to atomic gas.  There is more discussion on this topic in
\S\ref{s:star}.

One possible result of star formation is that the molecular gas becomes hotter.  One of the
diagnostics of high temperatures is the $^{12}$CO ($J=3 \to 2$)/($J=2 \to 1$) ratio. 
 NGC~5461 has a similar $^{12}$CO ($J=3 \to 2$)/($J=2 \to 1$)
ratio to the clouds in M33 that are associated with \hii\ regions \cite{55}.  However, the ratio
for NGC~5461 does not approach the value for the GMC NGC~604-2, which is, indeed, very high
compared to other ratios seen in M33 and other galaxies.  An explanation for the lower
 value of
the $^{12}$CO ($J=3 \to 2$)/($J=2 \to 1$) ratio in NGC~5461 compared to that found in other giant \hii\ regions is that
the clouds in NGC~5461 may have a mixture of warm and hot gas.  The dominant warm
 component could be similar in
temperature and density to the GMCs in M33 associated with \hii\ regions \cite{55}.

\section{Star Formation} \label{s:star}


The main motivation of this paper is to understand the
connection between the molecular gas in M101 and the
existence of the  bright stars that produce the giant \hii\ regions.  In the
Milky Way, the masses of GMCs that are affiliated with OB
associations are (1 -- 7)$\times 10^5$ \smass\ \cite{65}.  From this observational result, 
Williams and McKee 
\cite*{65} suggest that the probability a GMC will contain a
very bright massive star (O9.5) increases as the mass of the cloud
increases.  However, the GMCs in NGC~5461 have masses that are much smaller
than those required to form many bright O stars according to  Fig.\@ 7 of Williams 
and McKee \cite*{65}.  Perhaps the star-forming efficiency, and not
the mass of the clouds, is the key issue for regions with intense star formation.

\begin{figure}
\figurenum{4}
\figcaption{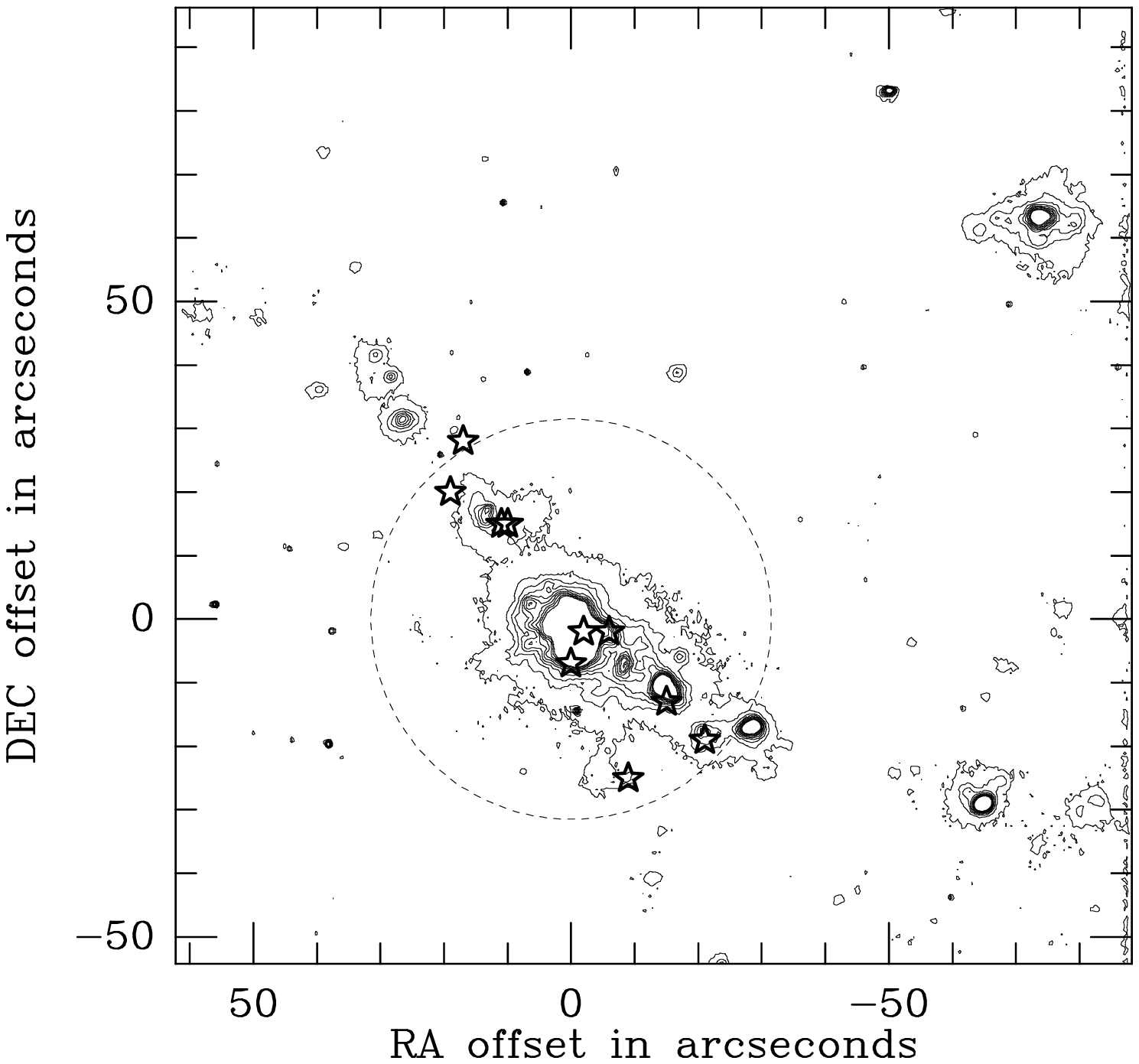}
\caption[Comparison of the CO with the H$\alpha$ map of
NGC~5461]{Comparison of the CO with the contour H$\alpha$ map of
NGC~5461.  The GMCs, which are denoted by stars,
surround the H$\alpha$ emission peaks.  The (0,0) position
is the center of the OVRO map, $\alpha = 14\timehr 01\timemin 55\fs6 $
and $\delta = +54\arcdeg  33\arcmin 31\farcs0$, and the circle
denotes the FWHM of the OVRO primary beam.\label{fig:halpha}}
\end{figure}

Intense star formation influences and is influenced by the surrounding 
ambient ionized
and atomic gas.  It is instructive to consider the correlation of molecular
gas with ionized and atomic gas.
One might expect that molecular and ionized hydrogen, which is
observed via the H$\alpha$ transition, should be associated with each
other.  In Figure~\ref{fig:halpha}, the CO peaks are compared with a
recent H$\alpha$ map that was kindly provided by Robert C. Kennicutt.
The CO clouds are denoted by stars to distinguish
them from the contours of the H$\alpha$ map.

\placefigure{fig:halpha}

The CO peaks  are typically displaced from the peak of the H$\alpha$ image by
$2\mbox{ -- }5''$ or, at the distance of M101, $70\mbox{ -- }180$ pc.
A possible reason, which was used to explain the same type of
anticorrelation in M51 \cite{33}, is that the presence of gas with
high surface density increases the extinction from the associated
dust.  The radiation from the ionized gas is obstructed, and therefore
the H$\alpha$ emission is anticorrelated with the molecular gas.

\begin{figure}
\figurenum{5}
\figcaption{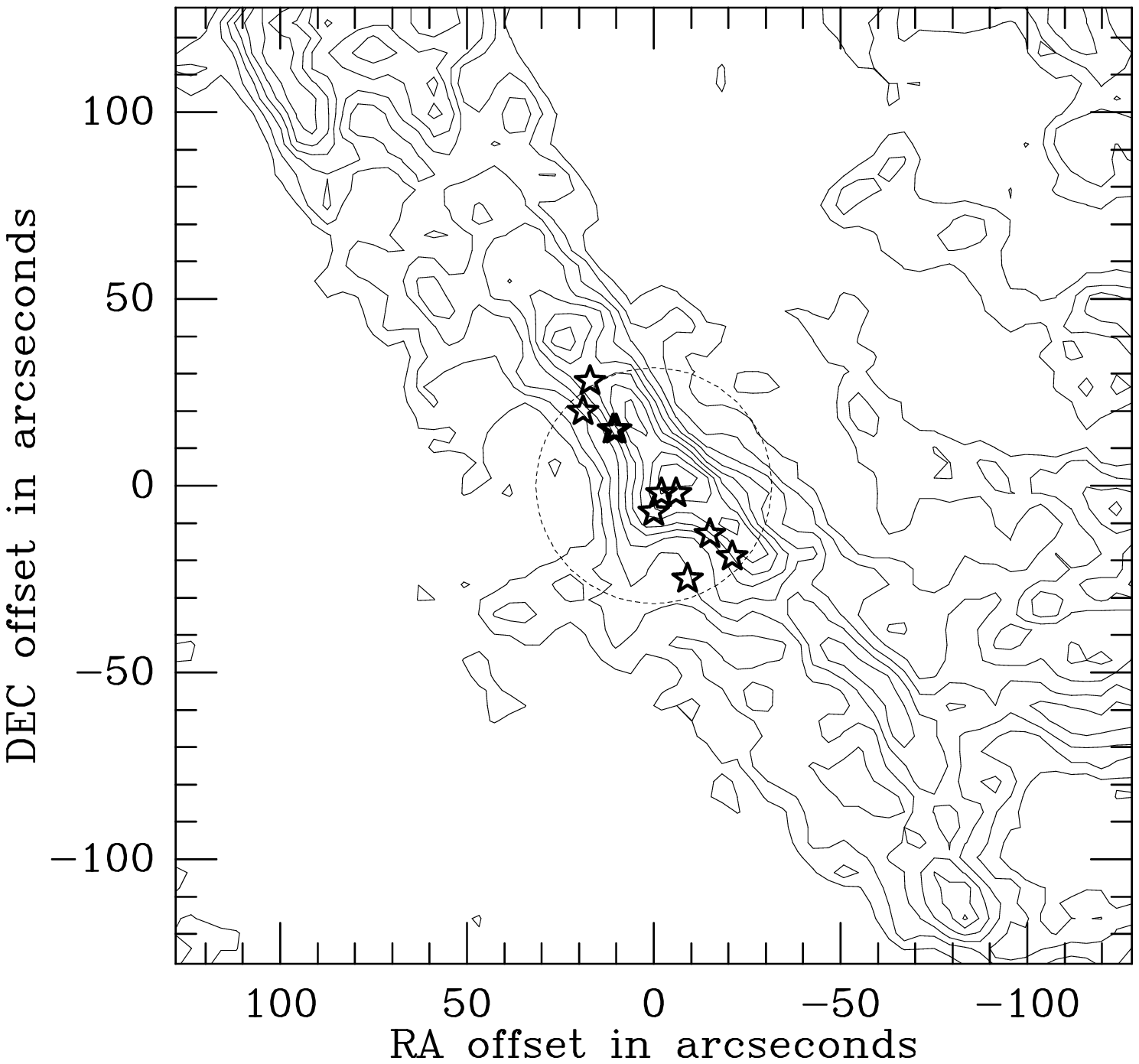}
\caption[Comparison of the CO with the \hi\ map of
NGC~5461]{Comparison of the CO clouds, which are denoted with stars,
with the atomic hydrogen contour map of NGC~5461.  The molecular
clouds in the CO map are offset by approximately 180 pc on average
from where the \hi\ emission peaks; however, the uncertainty in 
the positions is 140 pc.  The (0,0) position
is the center of the OVRO map, $\alpha = 14\timehr 01\timemin 55\fs6 $
and $\delta = +54\arcdeg  33\arcmin 31\farcs0$, and the circle
denotes the FWHM of the OVRO primary beam.\label{fig:hi}}
\end{figure}

Figure~\ref{fig:hi} presents the CO clouds in NGC~5461 superimposed on
a high resolution ($9''$) \hi\ contour map kindly provided
by Robert Braun. The GMCs are denoted with stars.  The
CO peaks are displaced compared to the peak of the \hi\ image by $3
\mbox{ -- }10''$ or, at the distance of M101, $100\mbox{ -- }360$ pc.
A possible explanation for this displacement is that the radiation
from the young massive stars in the \hii\ regions may have dissociated
the molecular gas to atomic gas; this explanation was used for M51
where a similar displacement was observed \cite{164}.  Unfortunately,
the uncertainty of the positions of the \hi\ map, $4''$, is fairly high;
it is possible that the observed offset is not significant.

\placefigure{fig:hi}

A large scale study of the atomic gas in M101 was conducted with 
$6''$ resolution \cite{63}.  Two of the several positions observed
coincide with the giant \HII\ regions NGC~5461 and NGC~5471.  The
atomic masses for the two regions are $5 \times 10^7 $ \smass\ and $ 4
\times 10^7$ \smass, respectively, which are a few times higher than the
corresponding molecular mass observed in similar beams.  This result
is consistent with the observed ratios of atomic to molecular gas in
a sample of 27 Sc galaxies \cite{57}.

Despite the fact that the masses of the atomic gas for NGC~5461 and
NGC~5471 are similar, their empirical \emph{molecular} masses are different by
a factor of 3.  Why does NGC~5461 have so much more
molecular gas than NGC~5471?  One possibility is that NGC~5461 has a more efficient mechanism to convert
atomic to molecular gas than NGC~5471 does.  The efficiency of the
conversion might be enhanced by large-scale gravitational effects that
increase gas interactions and remove loosely bound atomic gas
\cite{57}.  
Alternatively, it is possible that the star-formation mechanism in NGC~5471 has
had more time to dissociate the molecular gas (to form atomic gas) by the
intense radiation from the young stars \cite{164,38}.
Unfortunately, the available
data are not sufficient to find the relative ages of the giant \hii\ regions in
order to distinguish between  the two possibilities.

The sum of the masses of the molecular, 
ionized, and the atomic gas 
for both regions (Table~\ref{tab:efficiencies}) is in fair agreement
with the total gas mass calculated from the dust and the assumed 
 gas-to-dust ratio.  We
calculate the dust mass from the 60 $\mu$m-wavelength fluxes obtained
from the Infrared Astronomical Satellite (IRAS) Point Source Catalog
with the equation \cite{66}:
\begin{equation}
 M_{dust} = {4 a  \rho \over 3 Q_{em}} \, {S_{\nu} d^2 \over
B_{\nu}(T)}, 
\label{d_three}
\end{equation}
where $a$ is the radius of the dust grain, $\rho$ is the density of
the dust grain, $Q_{em}$ is the dust grain emission efficiency,
$S_{\nu}$ is the measured flux density, $d$ is the distance to the
dust (7.4 Mpc for the giant \hii\ regions in M101), and $B_{\nu}(T)$
is the Planck function. 
The flux at
60 $\mu$m for NGC~5461 is 9.65 Jy, and for NGC~5471 is 1.81 Jy
(NGC~5462 is not detected as a point source). For the
calculation, it has been assumed that $\rho=3$ g cm$^{-3}$, $Q_{em}/a
= 340$ cm$^{-1}$, and $T=30$ K \cite{66}, which are canonical values.  
With these assumptions, the
dust mass for NGC~5461 is $5 \times 10^5$ \smass\ while for NGC~5471
the dust mass is $9 \times 10^4$ \smass.  If the gas-to-dust ratio is
assumed to be a typical value of 600 for Sc~galaxies [Young and
Scoville \cite*{57} and references therein], then the total gas mass
in NGC~5461 is $3 \times 10^8$ \smass, and the total gas mass in
NGC~5471 is $5 \times 10^7$ \smass, which is in agreement with the sum of
all gas components in these regions (Table~\ref{tab:efficiencies}).  

\placetable{tab:efficiencies}



The star-formation efficiency is the ratio of the mass of stars formed
in the region to the sum of the stellar and molecular masses of the
 region \cite{176}.  Recently, the mass of stars  ($M >  2$ \smass) in a small
portion of NGC~5461 has been estimated from UV fluxes \cite{41}; however, to
compare this result with our data, we must 
scale the mass of stars to include all stars  down
 to 0.1 \smass, and we must scale the
flux to correct for the small aperture used ($1''$).  We estimate that
the first scale factor is 2.6 from Table 9 in Miller and Scalo \cite*{177}, and the 
second scale factor is 4 for both NGC~5461 and NGC~5471 from Figure 1 in 
Rosa and Benvenuti \cite*{41}.
The masses of the stars 
and the star formation efficiencies in NGC~5461 and NGC~5471
calculated in this manner are given in Table~\ref{tab:efficiencies}.  
The star forming efficiencies are slightly larger than
the efficiencies of a few percent (1\% $\mbox{ -- }$ 4\%) observed in
several clouds in the Milky Way such as the Taurus-Auriga, Orion A,
and Orion B clouds \cite{175}.  Thus, the GMCs in these giant \hii\ regions 
are somewhat more
 efficient in creating stars than molecular clouds in our Galaxy.  
 Perhaps the mass of
 the atomic gas should be included in the calculation of the star
 efficiency if the molecular gas has already been dissociated to \hi.
 However, the \hi\  has been found
 at great distances from the molecular cloud especially in the case of NGC~5471.
   It is not obvious that
 the \hi\ is directly connected with the star forming regions.


\section{Conclusions} \label{s:conclusions} %

In this paper, the molecular components of three giant \hii\ regions
 in the spiral galaxy M101 have been
investigated with new observations from two single dish telescopes
(JCMT and NRAO 12-meter) and from the OVRO millimeter array.

(1) NGC~5461 is the only \hii\ region with strong enough emission to be
detected with the OVRO millimeter array.  
The mass of the GMCs in NGC~5461 was calculated empirically and
from the virial theorem, and we found that the appropriate 
$X$ factor toward the giant \hii\ regions is 5 times smaller than
$X_{Gal}$. These data provide the first empirical demonstration that the
value of $X$ may decrease in regions with intense star formation. 
The corrected empirical masses of the
 large GMCs cover a range of $(2 \mbox{ -- } 8)\times 10^5$~\smass, which
 is comparable to that observed in the Galaxy.   
 Using this determined value of $X$, we calculate the
mass of the associations of GMCs in NGC~5461, NGC~5462, and NGC~5471.

(2)  The
molecular mass for the association of GMCs in NGC~5461 is calculated
to be $(15 \mbox{ -- } 40) \times 10^6$ \smass.  
The higher value in
the range comes from the LTE analysis, while the lower value comes from
the empirical method using NRAO data of NGC~5461.  The discrepancy would
be resolved if the $\eta_c$ was larger.  This efficiency probably is larger
because there could be more GMCs in NGC~5461 that were not detected.
The interferometric data provide a lower mass ($6 \times 10^6$ \smass)
because the flux from extended low-level emission is
typically lost in an interferometer.
The molecular mass toward NGC~5462 is estimated to be (2.2--3.5)$\times 10^6$ \smass.
CO emission was not detected toward NGC~5471; the upper limit to the mass
is $6.5 \times 10^6$ \smass.   

(3) The gas emitting the CO is very dense, so dense that it cannot
fill up more than 1\% of the volume without exceeding the GMC mass limits.  
  It is possible that the intense radiation field in the
vicinity of a giant \hii\ region partially dissociates or ionizes portions of the
molecular gas leaving small dense cores in large sparse envelopes.
 
(4)  The molecular mass of 
the NGC~5461 association of clouds is
accompanied by 1--2 times as much atomic mass.  The fairly large
ratio of atomic to molecular gas can be attributed either to slow
formation of GMCs from the original atomic clouds or to efficient
dissociation of the GMCs to atomic gas.   An even stronger presence of 
atomic hydrogen is observed in  the vicinity of
NGC~5471; in this region, the atomic mass is an order of magnitude larger
than molecular mass.  Perhaps NGC~5471 has not converted much of its atomic gas to
molecular gas yet, or the molecular gas has been dissociated to form atomic gas.  
For both NGC~5461 and NGC~5471, the total gas mass can be extrapolated
from the mass of the dust estimated from IRAS
data. The estimates of the total gas mass from the dust are
consistent with the sum of the molecular, ionized, and atomic masses.

(5)  The relatively normal masses observed toward
the clouds in NGC~5461 reinforce the hypothesis that giant \hii\
regions in M101 are so much brighter than the \hii\ regions in our
Galaxy because of the different properties of the natal clouds in
M101.  In particular,  
this paper suggests that the high star-formation efficiency of the
gas, and not the large mass
of the cloud, is the key
to the formation of giant \hii\ regions in M101.

\acknowledgments
The authors would like to thank Lorne Avery for his LVG code, Robert Braun for providing
his \hi\ maps of M101, David Frayer for helping with the NRAO figure,
Jeff Kenney for sharing his unpublished
CO map of M101, Robert Kennicutt for providing
the H$\alpha$ image of NGC~5461, and the staff at the JCMT.  
We are grateful to the
National Research Council of Canada for supporting observing
trips and to the Astronomy Department of the California Institute
of Technology. This work was partly supported by NSERC Canada.

\appendix


\section{Calculating the Radiation Temperature, $T_R$, from Observed Temperatures}

To calculate \label{etac} the radiation temperature, one needs $\eta_c$, the efficiency with which the antenna
diffraction pattern couples to the source \cite{68}: 
 \begin{equation}
   T_R =  {T_A^* \over \eta_c \; \eta_{fss}} = {T_R^* \over \eta_c}, 
    \label{twentytwo}
\end{equation}
where $\eta_{fss}$ is the efficiency of forward scattering and spillover.
The model used for the
integration takes into account both the size (as determined from the
high resolution OVRO data of NGC~5461) and the position of
each GMC with respect
to the center of the JCMT beam.  We found that the coupling efficiencies for NGC~5461
are $0.056 \pm 0.024$ and $0.037 \pm 0.016$ for the two transitions \COb\ and 
$J=3 \to 2$  respectively. 

One has to assume that the coupling efficiency for the
gas in NGC~5462 are similar to NGC~5461 because there were no
interferometric detections of clouds in NGC~5462.  This
assumption is probably not too bad because the values for the
molecular mass obtained with the LTE method are within a factor of two
or three from the values obtained from the empirical method, which
does not utilize the coupling efficiency (Table~\ref{tab:ansummary}).

\clearpage



\clearpage

\clearpage

\begin{deluxetable}{lrrccccc}
\small
\tablenum{1}
\tablecaption{Spectral line parameters for NGC~5461 with JCMT data\label{tab:NGC5461}}
\tablewidth{0pt}
\tablecolumns{8}
\tablehead{
\colhead{Source} & \colhead{$\Delta \alpha$} & \colhead{$\Delta \delta$} &
  \colhead{$\int{T_A^* dv}$} & \colhead{$V_{peak}$} & \colhead{$\Delta V$} &
  \colhead{$T_{peak}$} & \colhead{Integration} \\
\colhead{} & \colhead{($\arcsec$)} & \colhead{($\arcsec$)} &
  \colhead{(K km s$^{-1})$} &
  \colhead{(km s$^{-1})$} & \colhead{(km s$^{-1})$} & \colhead{(mK)} &
  \colhead{time (sec)}
}
\startdata
\cutinhead{\COb}
NGC~5461a &    0  &    0  &          $5.8  \pm 0.3$ & 276     & 24      &          $228 \pm    11$   &  3000 \nl
NGC~5461b &    7  &    7  &          $5.4  \pm 0.3$ & 277     & 24      &          $206 \pm    11$   &  1800 \nl
NGC~5461c &  $-7$ &  $-7$ &          $5.3  \pm 0.3$ & 273     & 24      &          $207 \pm    10$   &  1800 \nl
NGC~5461d &  $-7$ &    7  &          $4.7  \pm 0.4$ & 275     & 23      &          $181 \pm    11$   &  1800 \nl
NGC~5461e &    7  &  $-7$ &          $2.1  \pm 0.4$ & 273     & 24      &       $\phn83 \pm \phn9$   &  1800 \nl
NGC~5461f & $-14$ & $-14$ &          $5.6  \pm 0.3$ & 272     & 25      &          $212 \pm    11$   &  1650 \nl
NGC~5461g &   14  &   14  &          $3.5  \pm 0.4$ & 287     & 22      &          $158 \pm    12$   &  1800 \nl
NGC~5461h & $-14$ &    0  &          $5.9  \pm 0.3$ & 274     & 26      &          $217 \pm \phn9$   &  1800 \nl
NGC~5461i &   14  &    0  &          $1.7  \pm 0.4$ & 277     & 25      &       $\phn68 \pm    10$   &  1800 \nl
NGC~5461j &    0  &   14  &          $1.5  \pm 0.4$ & 280     & 17      &       $\phn79 \pm    10$   &  1800 \nl
NGC~5461k &    0  & $-14$ &          $5.3  \pm 0.3$ & 272     & 24      &          $202 \pm    11$   &  1800 \nl
\cutinhead{\COc}
NGC~5461a &    0  &   0   & \phm{$<$}$4.1  \pm 0.2$ & 277     & 23      & \phm{$<$}$180 \pm    12$   &  7200 \nl
NGC~5461b &    7  &   7   & \phm{$<$}$3.3  \pm 0.3$ & 281     & 20      & \phm{$<$}$140 \pm    15$   &  3000 \nl
NGC~5461c &  $-7$ & $-7$  & \phm{$<$}$5.7  \pm 0.2$ & 275     & 27      & \phm{$<$}$210 \pm    17$   &  3000 \nl
NGC~5461d &  $-7$ &   7   &       $<$$1.1  \pm 0.3$ & \nodata & \nodata &    $<$$\phn51 \pm    18$   &  3000 \nl
NGC~5461e &    7  & $-7$  &       $<$$1.1  \pm 0.3$ & \nodata & \nodata &    $<$$\phn53 \pm    24$   &  3000 \nl
\cutinhead{\tCOb}
NGC~5461a &    0  &   0   &         $0.49 \pm 0.05$ & 272     & 27.8    &           $17 \pm     4.0$ & 10840 \nl
\enddata
\tablecomments{The offsets are with respect to the center of NGC~5461:
	       $\alpha=14\timehr01\timemin55\fs6$ and $\delta = 54\arcdeg
	       33\arcmin31\farcs0$.  The quantity $\int{T_A^* dv}$ is the
	       integrated antenna temperature.  The full-width at half-maximum,
	       $\Delta V$, and the central velocity, $V_{peak}$, for which the
	       maximum antenna temperature value, $T_{peak}$, occurred were
	       calculated by fitting Gaussian lines to the spectra.}
\end{deluxetable}

\begin{deluxetable}{lrrccccc}
\small
\tablenum{2}
\tablecaption{Spectral line parameters for NGC~5462 with JCMT data\label{tab:NGC5462}}
\tablewidth{0pt}
\tablecolumns{8}
\tablehead{
\colhead{Source} & \colhead{$\Delta \alpha$} & \colhead{$\Delta \delta$} &
  \colhead{$\int{T_A^* dv}$} & \colhead{$V_{peak}$} & \colhead{$\Delta V$} &
  \colhead{$T_{peak}$} & \colhead{Integration} \\
\colhead{} & \colhead{(\arcsec)} & \colhead{(\arcsec)} &
  \colhead{(K km s$^{-1})$} & \colhead{(km s$^{-1})$} & \colhead{(km s$^{-1})$} &
  \colhead{(mK)} & \colhead{time (sec)}
}
\startdata
\cutinhead{\COb}
NGC~5462a &   0  &   0  &       $<$$0.10 \pm 0.10$ & \nodata & \nodata  &       $<$$\phn6 \pm 8$     &  1800 \nl
NGC~5462b &   0  &   6  & \phm{$<$}$0.51 \pm 0.10$ & 300 & 17.6 & \phm{$<$}$   28 \pm 8$     &  1800 \nl
NGC~5462c &   0  & $-6$ &       $<$$0.69 \pm 0.20$ & \nodata & \nodata  &       $<$$   19 \pm 7$     &  1800 \nl
NGC~5462d &   6  &   0  & \phm{$<$}$0.49 \pm 0.10$ & 304 & 17.3 & \phm{$<$}$   27 \pm 7$     &  1800 \nl
NGC~5462e & $-6$ &   0  & \phm{$<$}$0.57 \pm 0.10$ & 299 & 16.4 & \phm{$<$}$   32 \pm 7$     &  1800 \nl
NGC~5462f &   6  &   6  &       $<$$0.41 \pm 0.10$ & \nodata & \nodata  &       $<$$   18 \pm 7$     &  1800 \nl
NGC~5462g &   0  &  12  & \phm{$<$}$0.69 \pm 0.05$ & 292 & 12.0 & \phm{$<$}$   53 \pm 7$     &  1800 \nl
\cutinhead{\COc}
NGC~5462a &   0  &   0  &       $<$$0.45 \pm 0.4$  & \nodata & \nodata  & \phm{$<$}$   21 \pm \phn6$ &  4800 \nl
NGC~5462b &   0  &   6  & \phm{$<$}$0.20 \pm 0.4$  & 295 & 12   & \phm{$<$}$   18 \pm \phn6$ &  4800 \nl
NGC~5462d &   6  &   0  & \phm{$<$}$0.56 \pm 0.4$  & 296 & 23   & \phm{$<$}$   25 \pm \phn7$ &  4800 \nl
NGC~5462f &   6  &   6  &       $<$$0.66 \pm 0.2$  & \nodata & \nodata  &       $<$$   14 \pm \phn6$ &  6000 \nl
NGC~5462g &   0  &  12  & \phm{$<$}$1.13 \pm 0.2$  & 297 & 22   & \phm{$<$}$   50 \pm \phn6$ &  6000 \nl
NGC~5462h & $-6$ &   6  &       $<$$0.41 \pm 0.3$  & \nodata & \nodata  &       $<$$   15 \pm    12$ &  1200 \nl
NGC~5462n &   0  &  18  & \phm{$<$}$1.00 \pm 0.3$  & 300 & 27   & \phm{$<$}$   36 \pm    12$ &  1200 \nl
NGC~5462p &   6  &  12  &       $<$$0.94 \pm 0.3$  & \nodata & \nodata  &       $<$$   27 \pm    12$ &  1200 \nl
NGC~5462q & $-6$ &  12  &       $<$$0.49 \pm 0.3$  & \nodata & \nodata  & \phm{$<$}$   49 \pm    12$ &  1200 \nl
\cutinhead{\tCOb}
NGC~5462g &   0  &  12  &          $0.10 \pm 0.01$ & 295 &  6   &             $10 \pm 2.5$   & 15600 \nl
\enddata
\tablecomments{The offsets are with respect to the center of NGC~5462:
	$\alpha=14\timehr02\timemin07\fs6$ and $\delta = 54\arcdeg36\arcmin
	17\farcs4$.  The quantity $\int{T_A^* dv}$ is the integrated antenna
	temperature.  The full-width at half-maximum, $\Delta V$, and the
	central velocity, $V_{peak}$, for which the maximum antenna temperature
	value, $T_{peak}$, occurred were calculated by fitting Gaussian lines to
	the spectra.}
\end{deluxetable}

\begin{deluxetable}{ccccc}
\tablenum{3}
\tablecaption{\COa\ spectral line parameters for the NRAO data\label{tab:nrao}}
\tablewidth{0pt}
\tablehead{
\colhead{Source} & \colhead{$\int{T_R^* \, dv}$} & \colhead{$V_{peak}$} &
  \colhead{$\Delta V$ (FWHM)} & \colhead{$T_{R_{peak}}^*$} \\
\colhead{} &  \colhead{(K km s$^{-1}$)} & \colhead{(km s$^{-1}$)} &
  \colhead{(km s$^{-1}$)} & \colhead{(mK)}
}
\startdata
NGC~5461 & \phm{$<$}$2.50 \pm 0.48$ & 274     & 29      & \phm{$<$}$78 \pm \phn8$ \nl 
NGC~5462 & \phm{$<$}$0.38 \pm 0.11$ & 296     & 30      & \phm{$<$}$12 \pm \phn2$ \nl
NGC~5471 &       $<$$0.54 \pm 0.20$ & \nodata & \nodata &       $<$$12 \pm    12$ \nl
\enddata
\tablecomments{The quantity $\int{T_R^* dv}$ is the integrated corrected
radiation temperature.  The full-width at half-maximum, $\Delta V$,
and the central velocity, $V_{peak}$, for which the maximum corrected radiation
temperature value, $T_{R_{peak}}^*$, occurred were calculated by fitting
Gaussian lines to the spectra.  The coordinates of the center of NGC~5471 are 
$\alpha=14\timehr02\timemin43\fs5$ and $\delta = 54\arcdeg38\arcmin
	09\farcs0$.}
\end{deluxetable}

\begin{deluxetable}{cccccccc}
\tablenum{4}
\tablecaption{Measured properties of molecular
  clouds in NGC~5461\label{tab:ovro_gma}}
\tablewidth{0pt}
\tablehead{
\colhead{} &  \colhead{Central} & \colhead{} & \colhead{Offset} & \colhead{} &
  \colhead{Total} & \colhead{} \\
\colhead{Cloud} & \colhead{Velocity} & \colhead{Dimensions} &
  \colhead{Position} & \colhead{$\Delta v$} & \colhead{integrated flux} &
  \colhead{$T_B$} \\
\colhead{} & \colhead{(km s$^{-1})$} & \colhead{(pc$\times$pc)} &
  \colhead{($\arcsec,\arcsec$)} & \colhead{(km s$^{-1})$} &
  \colhead{(Jy km s$^{-1}$)} & \colhead{(K)}
}
\startdata
1  & 263 & $\phn99\times 90$ & (\phs19,\phs20)      & \phn7.8 & $ 2.6 \pm 0.9$ & 7.5 \nl
2  & 270 & $   108\times 90$ & (\phn$-$2,\phn$-$2)  &    10.4 & $ 4.4 \pm 1.0$ & 5.3 \nl
3  & 270 & $   180\times 81$ & (\phs\phn0,\phn$-$7) & \phn5.2 & $ 1.9 \pm 0.5$ & 5.5 \nl
4  & 271 & $\phn99\times 73$ & ($-$21,$-$19)        & \phn7.8 & $ 3.0 \pm 0.9$ & 8.1 \nl
5  & 274 & $\phn94\times 73$ & (\phn$-$6,\phn$-$2)  & \phn7.8 & $ 1.2 \pm 0.3$ & 4.5 \nl
6  & 276 & $   140\times 90$ & ($-$15,$-$13)        & \phn7.8 & $ 4.7 \pm 1.2$ & 6.1 \nl
7  & 282 & $   126\times 81$ & (\phs10,\phs15)      & \phn7.8 & $ 3.3 \pm 0.7$ & 6.6 \nl
8  & 285 & $\phn99\times 73$ & (\phs17,\phs28)      & \phn5.2 & $ 2.1 \pm 0.6$ & 7.7 \nl
9  & 285 & $   108\times 90$ & (\phn$-$9,$-$25)     & \phn5.2 & $ 2.2 \pm 0.6$ & 6.0 \nl
10 & 298 & $   180\times 90$ & (\phs11,\phs15)      & \phn5.2 & $ 2.8 \pm 0.9$ & 5.7 \nl
\enddata
\tablecomments{The dimensions are not deconvolved from the beam.  The offset
positions are with respect to the center of NGC~5461:  $\alpha=14\timehr
01\timemin55\fs6$ and $\delta = 54\arcdeg 33\arcmin 31\farcs0$. The integrated
fluxes and brightness temperatures have been corrected for the primary beam falloff in sensitivity.}
\end{deluxetable}

\begin{deluxetable}{cccccc}
\tablenum{5}
\tablecaption{Sizes and masses of GMCs in NGC~5461
\label{tab:sizeline}}
\tablewidth{0pt}
\tablehead{
\colhead{Molecular} & \colhead{$\Delta V_{obs}$} & 
  \colhead{$D_{exp}$} & \colhead{$M_{vir}$} & \colhead{$M_{mol}$} &\colhead{Corrected $M_{mol}$} \\
\colhead{Cloud} & \colhead{(km s$^{-1}$)} & 
  \colhead{(pc)} & \colhead{($10^5$  \smass)} & \colhead{($10^5$ \smass)} & \colhead{($10^5$ \smass)}
}
\startdata
 1  & \phn7.8 & 42 & $<$$\phn6 \pm 2$ & $46 \pm    16$ & $5 \pm 2$ \nl
 2  &    10.4 & 75 & $<$$   11 \pm 3$ & $78 \pm    17$ & $8 \pm 2$ \nl
 3  & \phn5.2 & 19 & $<$$\phn3 \pm 1$ & $34 \pm \phn9$ & $3 \pm 1$ \nl
 4  & \phn7.8 & 42 & $<$$\phn6 \pm 2$ & $54 \pm    16$ & $5 \pm 2$ \nl
 5  & \phn7.8 & 42 & $<$$\phn6 \pm 2$ & $22 \pm \phn5$ & $2 \pm 1$ \nl
 6  & \phn7.8 & 42 & $<$$\phn6 \pm 2$ & $85 \pm    22$ & $8 \pm 2$ \nl
 7  & \phn7.8 & 42 & $<$$\phn6 \pm 2$ & $59 \pm    13$ & $6 \pm 1$ \nl
 8  & \phn5.2 & 19 & $<$$\phn3 \pm 1$ & $38 \pm    11$ & $4 \pm 1$ \nl
 9  & \phn5.2 & 19 & $<$$\phn3 \pm 1$ & $40 \pm    11$ & $4 \pm 1$ \nl
 10 & \phn5.2 & 19 & $<$$\phn3 \pm 1$ & $50 \pm    16$ & $5 \pm 2$ \nl
\enddata
\tablecomments{The velocity dispersion, $\Delta V_{obs}$,
 of the ten clouds in NGC~5461 is presented along
with the expected size, $D_{exp}$, given the M33 size:line-width relationship. 
  In addition,
we present the virial masses and the empirical masses of the GMCs in NGC~5461.
The uncertainties in the empirical masses take into account only the uncertainty
in the measurement of the integrated flux.  The corrected empirical masses  are
calculated for the new reduced value of $X$ toward NGC~5461.}
\end{deluxetable}

\begin{deluxetable}{lc}
\tablenum{6}
\tablecolumns{2}
\tablecaption{Summary of masses obtained in the analysis 
\label{tab:ansummary}}
\tablewidth{0pt}
\tablehead{
\colhead{Method} & \colhead{Mass ($\times 10^5$ \smass)}
}
\startdata
\cutinhead{NGC~5461}
LTE\tablenotemark{a}              & $  400 \; [350\mbox{ -- }500]$ \nl
Empirical (NRAO 12-meter)         & $ 150 \pm 30$               \nl
Empirical (OVRO) & $  60 \pm 23$               \nl
\sidehead{Average cloud in NGC~5461:}
Empirical (OVRO) & $   5$                    \nl

\cutinhead{NGC~5462}
LTE\tablenotemark{a}              & $ >35$ \nl
Empirical (NRAO 12-meter)         & $  22 \pm 7$               \nl
Empirical (OVRO) & $  <30$                      \nl
\cutinhead{NGC~5471}
Empirical (NRAO 12-meter)         & $<65 \pm 24$               \nl
\enddata
\tablenotetext{a}{Method depends on the value of the coupling
  efficiency, $\eta_c$.}
\end{deluxetable}

\begin{deluxetable}{cccc}
\tablenum{7}
\tablecaption{Radiation temperature ratios\label{tab:ratios}}
\tablewidth{0pt}
\tablehead{
\colhead{Source} &
  \colhead{\rule[-3ex]{0pt}{6ex}${\displaystyle \frac{^{12}{\rm CO}\; J=3\to2}{^{12}{\rm CO}\; J=2\to1}}$} & 
  \colhead{${\displaystyle \frac{^{12}{\rm CO}\; J=1\to0}{^{12}{\rm CO}\; J=2\to1}}$} & 
  \colhead{${\displaystyle \frac{^{13}{\rm CO}\; J=2\to1}{^{12}{\rm CO}\; J=2\to1}}$}
}
\startdata
NGC~5461 & $0.80 \pm 0.04$ & $0.56 \pm 0.11$ & $0.147 \pm 0.015$ \nl
NGC~5462 & $1.14 \pm 0.22$ & \nodata         & $0.135 \pm 0.030$ \nl
\enddata
\tablecomments{To obtain the observed ratios, we
convolved the high-resolution data to the same resolution as the low-resolution data.
The lines were divided channel-by-channel and averaged over all channels with a
signal-to-noise ratio greater than 2.}
\end{deluxetable}

\begin{deluxetable}{ccccc}
\tablenum{8}
\tablecaption{Masses and star forming efficiencies
\label{tab:efficiencies}}
\tablewidth{0pt}
\tablehead{
\colhead{Name} & \colhead{Ionized mass} & \colhead{Stellar mass} & \colhead{Molecular Mass} &
   \colhead{Star formation efficiencies} \\
\colhead{}  & \colhead{($\times 10^6$ \smass)}   & \colhead{($\times 10^6$ \smass)} &
\colhead{($\times 10^6$ \smass)} & \colhead{\%}
}
\startdata
NGC 5461 & 30 & 1\phd\phn & \phm{$<$}15\phd\phn & \phm{$>$}\phn6 \nl
NGC 5471 & 14 & 0.8       &    $<$\phn6.5       &          $>$11 \nl
\enddata
\tablecomments{The mass of the ionized gas is from Israel et al.\@  \cite*{11},
and the stellar mass is estimated from Rosa and Benvenuti \cite*{41} after
correcting to include low mass stars ($M>0.1$ \smass) and the proper aperture
size.}
\end{deluxetable}

\end{document}